\documentclass[%
reprint,
 amsmath,amssymb,
 aps,
prb,
floatfix
]{revtex4-2}

\usepackage{graphicx}% Include figure files
\usepackage{dcolumn}% Align table columns on decimal point
\usepackage{bm}% bold math
\usepackage{hyperref}% add hypertext capabilities
\usepackage{braket} % to use bra-ket notation

\begin{document}

% \preprint{APS/123-QED}

\title{Efficient calculation of phonon dynamics through a low-rank solution of the Boltzmann equation}% Force line breaks with \\

\author{Nikhil Malviya}
\author{Navaneetha K. Ravichandran}
\email{navaneeth@iisc.ac.in}
\affiliation{%
 Department of Mechanical Engineering, Indian Institute of Science, Bangalore 560012, India
 }%

\date{\today}% It is always \today, today,

\begin{abstract}
Exotic nondiffusive heat transfer regimes such as the second sound, where heat propagates as a damped wave at speeds comparable to those of mechanical disturbances, often occur at cryogenic temperatures (T) and nanosecond timescales in semiconductors. First-principles prediction of such rapid, low-T phonon dynamics requires finely-resolved temporal tracking of large, dense, and coupled linear phonon dynamical systems arising from the governing linearized Peierls-Boltzmann equation (LPBE). Here, we uncover a rigorous low-rank representation of these linear dynamical systems, derived from the spectral properties of the phonon collision matrix, that accelerates the first-principles prediction of phonon dynamics by a factor of over a million without compromising on the computational accuracy. By employing this low-rank representation of the LPBE, we predict strong amplification of the wave-like second sound regime upon isotopic enrichment in diamond - a finding that would have otherwise been computationally intractable using the conventional brute-force approaches. Our framework enables a rapid and accurate discovery of the conditions under which wave-like heat flow can be realized in common semiconductors.
% Exotic nondiffusive heat transfer regimes such as the second sound, where heat propagates as a damped wave at speeds comparable to that of sound, occur at cryogenic temperatures (T) and nanosecond timescales in semiconductors. First-principles prediction of such rapid, low-T phonon dynamics requires finely-resolved temporal tracking of dense, coupled, linear phonon dynamical systems arising from the linearized Peierls-Boltzmann equation (LPBE). Here, we uncover a low-rank representation of the LPBE, derived from the spectral properties of the phonon collision matrix, that drives a million-fold acceleration of the first-principles prediction of phonon dynamics, while preserving accuracy. Using this low-rank  LPBE framework, we predict strong amplification of the second sound signature upon isotopic enrichment in diamond - a finding that would have otherwise been computationally intractable using the conventional brute-force approaches. Our framework enables a rapid and accurate discovery of the conditions under which wave-like heat flow can be realized in common semiconductors.
\end{abstract}

\maketitle

\clearpage

\section{Introduction} \label{sec:introduction}
Hydrodynamic heat transport, a type of nondiffusive thermal transport, has recently captured the attention of the research community, as it allows rapid heat propagation at rates comparable to those of mechanical disturbances, thus challenging the conventional picture of heat flow within the framework of the Fourier's law. Under steady-state conditions, the hydrodynamic heat flow manifests itself as the Poiseuille flow regime for phonons, which resembles a pressure-driven viscous fluid flow in a pipe, and has been experimentally observed in several materials such as bismuth~\cite{markov_hydrodynamic_2018}, strontium titanate~\cite{martelli_thermal_2018}, black phosphorus~\cite{machida_observation_2018}, and graphite~\cite{ding_phonon_2018, machida_phonon_2020, li_reexamination_2022, guo_size_2021, huang_mapping_2022, guo_basal-plane_2023, huang_observation_2023, huang_graphite_2024} at temperatures of a few 10's of Kelvin. Similarly, the transient manifestation of the hydrodynamic heat flow regime called the second sound, where the imposed thermal disturbance travels as a damped temperature wave, has been observed in large single crystals of high purity sodium fluoride (NaF) under cryogenic conditions several decades ago~\cite{jackson_second_1970, jackson_thermal_1971, mcnelly_heat_1970, narayanamurti_observation_1972}. Recent experimental observations of the second sound in graphite at temperatures approaching 200 K~\cite{huberman_observation_2019, ding_observation_2022, jeong_transient_2021} make it possible to propose new paradigm solutions to conventional problems of efficient heat dissipation in industry-scale and consumer electronics, and open up new possibilities for thermal waveguides, cloaking, and shielding applications~\cite{shi_nonresistive_2019, lee_hydrodynamic_2020, chen_non-fourier_2021}.\\

In such nonmetallic crystals, heat is carried by phonons, which are quantized lattice vibrations of the crystal. Phonon transport through these materials is governed by the linearized Peierls-Boltzmann equation (LPBE)~\cite{peierls_zur_1929}, a set of linear, multidimensional, coupled partial differential equations. The steady-state solution of the LPBE, with the phonon collision matrix derived from first principles, has been extremely successful in describing the observed thermal conductivities ($\kappa$) of several naturally occurring semiconductors over the past two decades without any adjustable parameters~\cite{broido_lattice_2005, esfarjani_heat_2011, jain_strongly_2015, feng_four-phonon_2017, shulumba_lattice_2017, carrete_almabte_2017, ravichandran_phonon-phonon_2020, farris_microscopic_2024}. Furthermore, the predictive power of this first-principles framework has also driven the recent discovery of new ultrahigh-$\kappa$ materials that are not present in nature~\cite{li_high_2018, kang_experimental_2018, tian_unusual_2018, chen_ultrahigh_2020} and their unusual thermal properties~\cite{ravichandran_non-monotonic_2019, ravichandran_exposing_2021, li_anomalous_2022}.\\

In stark contrast, predictive calculations of the ultrafast transient phonon dynamics from first principles are rare~\cite{hua_space-time_2020, chiloyan_greens_2021}, due to the extreme demand on the computational resources placed by these calculations. For example, in simulating the second sound regime, which manifests as fast temperature oscillations superimposed on a slow background thermal decay in the experimental signals~\cite{jackson_second_1970, jackson_thermal_1971, mcnelly_heat_1970, narayanamurti_observation_1972, huberman_observation_2019, ding_observation_2022, jeong_transient_2021}, contributions to the complete first-principles solution of the LPBE from a broad temporal frequency spectrum with fine frequency resolution must be computed. Furthermore, under cryogenic conditions, fine numerical discretization of the phonon Brillouin zone is necessary to resolve the weak, but essential, momentum-dissipating Umklapp scattering events that the phonons with small wave-vector undergo (see section 69 in ref.~\cite{pitaevskii_physical_2012}). This additional requirement dramatically increases the computational cost to solve the LPBE for each temporal frequency considered in the transient calculations. Several past works overcome these computational hurdles by resorting to approximate models~\cite{hua_transport_2014, hua_analytical_2014, hua_heat_2018, zhang_emergence_2022, ezzahri_thermal_2022} which may not be applicable for ultrahigh-$\kappa$ materials or materials under cryogenic conditions~\cite{guyer_solution_1966, malviya_failure_2023}.\\

Here, we overcome these limitations by uncovering an optimal low-rank representation of the LPBE that enables a million-fold acceleration of its full first-principles solution, without compromising on the computational accuracy. This low-rank representation emerges from the realization that while phonons throughout the Brillouin zone scatter among themselves to resist heat flow even under cryogenic conditions~\cite{pitaevskii_physical_2012}, only a small set of special collections of these phonons, embodied by the eigenmodes of the phonon collision matrix with the smallest eigenvalues, contribute to thermal transport. We deploy this low-rank, first-principles LPBE solution to efficiently track the transient dynamics of an externally-imposed thermal disturbance in diamond – the highest $\kappa$ material known to date, at a millionth of the computational cost compared to conventional methods~\cite{hua_space-time_2020, chiloyan_greens_2021}, and expose strong amplification of the second sound signature in diamond at 100 K upon isotopic enrichment. Additionally, we leverage the symmetries of the phonon collision matrix derived from the space-group of the crystal to obtain the transient second sound dynamics in isotopically enriched diamond at 100 K in a few 100's of CPU hours, in spite of the stringent requirement of ultrafine temporal resolution as well as the highly-refined numerical discretization of Brillouin zone for the LPBE solution under these conditions. Our accelerated low-rank LPBE solution from first principles will enable a rapid search of conditions in materials beyond graphite, where strong hydrodynamic signatures can be experimentally realized.\\

\section{Steady-state and transient solutions of the LPBE} \label{sec:transient_solution_of_lpbe}
The LPBE describes the equilibration of the linearized nonequilibrium phonon distribution function, $f'_{\lambda} \left( \mathbf{x}, t \right)$, for the phonon mode $\lambda \equiv \left[ \mathbf{q}, j \right]$ with wave vector $\mathbf{q}$ and polarization $j$ at position $\mathbf{x}$ and time $t$~\cite{peierls_zur_1929}. The LPBE is given by:
\begin{align}
    \frac{\partial f'_{\lambda}}{\partial t} + \mathbf{v}_{\lambda} \cdot \nabla f'_{\lambda} = & - \sum_{\lambda_{1}} \Omega_{\lambda \lambda_{1}} f'_{\lambda_{1}} + \dot{\mathcal{H}}_{\lambda}
    \label{eq:lpbe_time_spatial_domain}
\end{align}
where $\mathbf{v}_{\lambda}$ is the phonon group velocity, $\Omega$ is the collision matrix as described in the Appendix~\ref{sec:phonon_collision_matrix}, and $\dot{\mathcal{H}}_{\lambda} = \dot{\mathcal{H}}_{\lambda} \left( \mathbf{x}, t \right) = \dot{\mathcal{Q}}_{\lambda} \left( \mathbf{x}, t \right)/( \hbar \omega_{\lambda} \sqrt{f_{\lambda}^{0} \left( f_{\lambda}^{0} + 1 \right)})$ is the source term corresponding to the phonon-specific rate of energy input $\dot{\mathcal{Q}}_{\lambda}$, with $\omega_{\lambda}$ being the phonon frequency and $f_{\lambda}^{0}$ being the equilibrium Bose-Einstein distribution. Upon solving Eq.~\ref{eq:lpbe_time_spatial_domain} for $f'_{\lambda} \left( \mathbf{x}, t \right)$, the temperature deviation, $\Delta T \left( \mathbf{x}, t \right)$, is obtained by relating the macroscopic crystal energy $C_{0} \Delta T$ to the energy of the phonon system: $(1/V)\sum_\lambda \hbar\omega_\lambda \sqrt{f_{\lambda}^{0} \left( f_{\lambda}^{0} + 1 \right)} f'_\lambda$, where $C_{0}$ is the volumetric heat capacity and $V$ is the crystal volume (see Supplementary section~S1).\\

Noting that the collision matrix $[\Omega_{\lambda \lambda_{1}}]$ is symmetric~\cite{hardy_phonon_1970}, its eigenvectors $\{ \mathfrak{e}^m_\lambda\}$ form a complete orthonormal basis, thus allowing the expansion of $f'_{\lambda}$ and $\dot{\mathcal{H}_{\lambda}}$ in terms of $\{ \mathfrak{e}_{\lambda}^{m} \}$ as:
\begin{equation*}
    f'_{\lambda} = \sum_{m} \vartheta^{m}  \mathfrak{e}_{\lambda}^{m}
    \text{, and }
    \dot{\mathcal{H}}_{\lambda} = \sum_{m} h^{m} \mathfrak{e}_{\lambda}^{m}
    % \label{eq:linear_combination_of_e}
\end{equation*}
were $m$ is the index for eigenvector $\mathfrak{e}^{m}$ of $\Omega$ with eigenvalue $\sigma^{m}$, and the space and time dependencies of $f'_{\lambda}$ and $\dot{\mathcal{H}_{\lambda}}$ are absorbed into the coefficients of the expansions - $\vartheta^{m}$ and $h^{m}$ respectively. The collision matrix $\Omega$ is known to have a null eigenvector corresponding to thermal equilibrium at a temperature $T_0$ given by: $\mathfrak{e}_{\lambda}^{m = 0} \left( T_{0} \right) = \sqrt{f_{\lambda}^{0} \left( f_{\lambda}^{0} + 1 \right)} \left( \hbar \omega_{\lambda} \right)/\sqrt{V k_{B} T_{0}^{2} C_{0}}$ with $\sigma^{0} = 0$~\cite{pitaevskii_physical_2012, krumhansl_thermal_1965, guyer_solution_1966, guyer_thermal_1966}.\\

Following the works of Hardy~\cite{hardy_phonon_1970} and Cepellotti \& Marzari~\cite{cepellotti_thermal_2016}, we use this eigenmode expansion to solve the steady-state LPBE and obtain the $\kappa$ as:
\begin{equation}
    \kappa_{ij} = C_{0} \sum_{m} \frac{\mathcal{V}^{0 m}_{i}}{\sqrt{\sigma^{m}}} \frac{\mathcal{V}^{0 m}_{j}}{\sqrt{\sigma^{m}}}
    \label{eq:eigenmode_kappa}
\end{equation}
where the $\mathcal{V}^{0m}_{i} = \sum_{\lambda} \mathfrak{e}_{\lambda}^{0} v_{i \lambda} \mathfrak{e}_{\lambda}^{m}$ is the velocity of the $m^{\mathrm{th}}$ eigenvector of $\Omega$ with a mode-specific thermal diffusivity $\left( \mathcal{V}^{0m}_{i} \right)^{2}/\sigma^{m}$ in the $i^{\mathrm{th}}$ Cartesian direction.\\

To study transient phonon dynamics, we begin by adopting the solution developed by Hua \& Lindsay~\cite{hua_space-time_2020} for the time-dependent LPBE (eq.~\ref{eq:lpbe_time_spatial_domain}). As shown in the Supplementary section~S1, the Fourier-transformed transient temperature response, $\Delta \tilde{T}$, is obtained as:
\begin{equation}
    \Delta \tilde{T} \left( \mathbf{\xi}, \eta \right) = \frac{\tilde{h}^{0}/\zeta}{\displaystyle - i \eta + \sum_{m, n > 0} \mathcal{P}^{m n} \left( \mathbf{\xi}, \eta \right) \cfrac{\mathcal{V}^{ 0 m} \cdot \mathbf{\xi}}{\sqrt{\sigma^{m}}} \cfrac{\mathcal{V}^{ n 0} \cdot \mathbf{\xi}}{\sqrt{\sigma^{n}}}}
    \label{eq:f_delta_T}
\end{equation}
where, the tildes over the variables represent their Fourier transforms, with $\mathbf{\xi}$ and $\eta$ being the spatial wave-vector and temporal frequency respectively, $\zeta = \sqrt{V C_{0}/k_{B} T_{0}^{2}}$, and $\mathcal{P}^{m n} \left( \mathbf{\xi}, \eta \right) = \sum_{j} \mathfrak{p}_{j}^{n} \mathfrak{p}_{j}^{m}/\left( 1 - i \rho_{j} \right)$, with $\mathfrak{p}_{j}$ and $\rho_{j}$ being the $j^{th}$ eigenvector and eigenvalue of an intermediate matrix $\Psi \left( \mathbf{\xi}, \eta \right)$ whose components are:
\begin{equation}
    \Psi^{n m} \left( \mathbf{\xi}, \eta \right) = \eta \frac{\Delta_{n m}}{\sigma^{n}} + \frac{\mathcal{V}^{n m}}{\sqrt{\sigma^{n} \sigma^{m}}} \cdot \mathbf{\xi}
    \label{eq:psi_matrix_element}
\end{equation}
where $\mathcal{V}^{nm}_{i} = \sum_{\lambda} \mathfrak{e}_{\lambda}^{n} v_{i \lambda} \mathfrak{e}_{\lambda}^{m}$. Finally, the spatio-temporal temperature dynamics of the phonon system is obtained by performing an inverse Fourier transform of $\Delta \tilde{T} \left( \mathbf{\xi}, \eta \right)$ with respect to $\eta$ and $\xi$.

\section{Rank reduction of the LPBE driven by the spectral properties of the collision matrix} \label{sec:results}
Hua \& Lindsay~\cite{hua_space-time_2020} used the above-described approach to obtain $\Delta\bar{T}$ for an impulsive heat source $\dot{\mathcal{H}}\left(\mathbf{x}, t\right)$ from first principles. However, this approach encounters two major computational hurdles from a practical stand-point. First, to track the fast oscillatory signatures of the second sound regime at timescales of a few nanoseconds, superimposed on a slow background thermal decay at timescales of a few 100 nanoseconds, as observed in experiments on NaF and graphite~\cite{jackson_second_1970, jackson_thermal_1971, mcnelly_heat_1970, narayanamurti_observation_1972, huberman_observation_2019, ding_observation_2022, jeong_transient_2021}, the frequency-domain solution $\Delta \tilde{T} \left( \mathbf{\xi}, \eta \right)$ must be computed spanning over a large range of temporal frequencies $\eta$, while maintaining a sufficiently fine resolution at low frequencies, for every spatial wave-vector $\xi$. This requirement translates into several computationally-expensive diagonalizations of the intermediate matrix $\Psi$, which is of the same size as the phonon collision matrix $\Omega$, for every $\{ \mathbf{\xi}, \eta \}$ pair, independently. In contrast, the calculation of the steady-state $\kappa$ effectively requires a single diagonalization of $\Omega$~\cite{fugallo_ab_2013, cepellotti_thermal_2016}. Second, even though strong second sound signatures are expected in materials under conditions where momentum-conserving Normal collisions dominate the scattering of the thermally-populated small wave-vector phonons, a small number of momentum-dissipating Umklapp events involving large wave-vector phonons are necessary for achieving a convergent steady-state $\kappa$~\cite{zhang_cryogenic_2024} and for fixing the phonon drift velocity to close the transient problem (see section 69 in ref.~\cite{pitaevskii_physical_2012}). Hence, a fine discretization of the phonon Brillouin zone is required to accurately account for the phonon scattering events across the entire Brillouin zone, particularly for ultrahigh-$\kappa$ materials like diamond at cryogenic temperatures, resulting in a collision matrix of about 70 billion elements, as detailed in the Supplementary section~S3. This additional complexity makes the diagonalization of the intermediate matrix $\Psi$ for each $\{ \mathbf{\xi}, \eta \}$ pair, prohibitively expensive.\\

Here, we uncover a strategy for accelerating the computation by exploiting important features of the phonon collision matrix $\Omega$, which have never been used to accelerate these calculations in the past. As pointed out by Hardy~\cite{hardy_phonon_1970}, the collision matrix is even with respect to its indices, i.e., $\Omega_{\lambda\lambda_{1}} = \Omega_{\left(-\lambda\right)\left(-\lambda_{1}\right)}$, where $\left(-\lambda\right) \equiv \left[-\mathbf{q}, j\right]$. Hence, the eigenmodes $\mathfrak{e}^m_\lambda$ can be chosen to be even or odd with respect to the phonon indices $\lambda$, i.e., $\mathfrak{e}^m_\lambda = \pm\mathfrak{e}^m_{-\lambda}$. The equilibrium null eigenmode $\mathfrak{e}^0_\lambda$ is even with respect to its indices; consequently, only the odd eigenmodes have a non-zero velocity $\mathcal{V}^{0m}$ (since the phonon velocity $v_{i\lambda}$ is odd with respect to $\lambda$) and thus contribute to $\kappa$, as per Eq.~\ref{eq:eigenmode_kappa}.\\

\begin{figure}[h]
    \centering
    \includegraphics[width=1.0\linewidth]{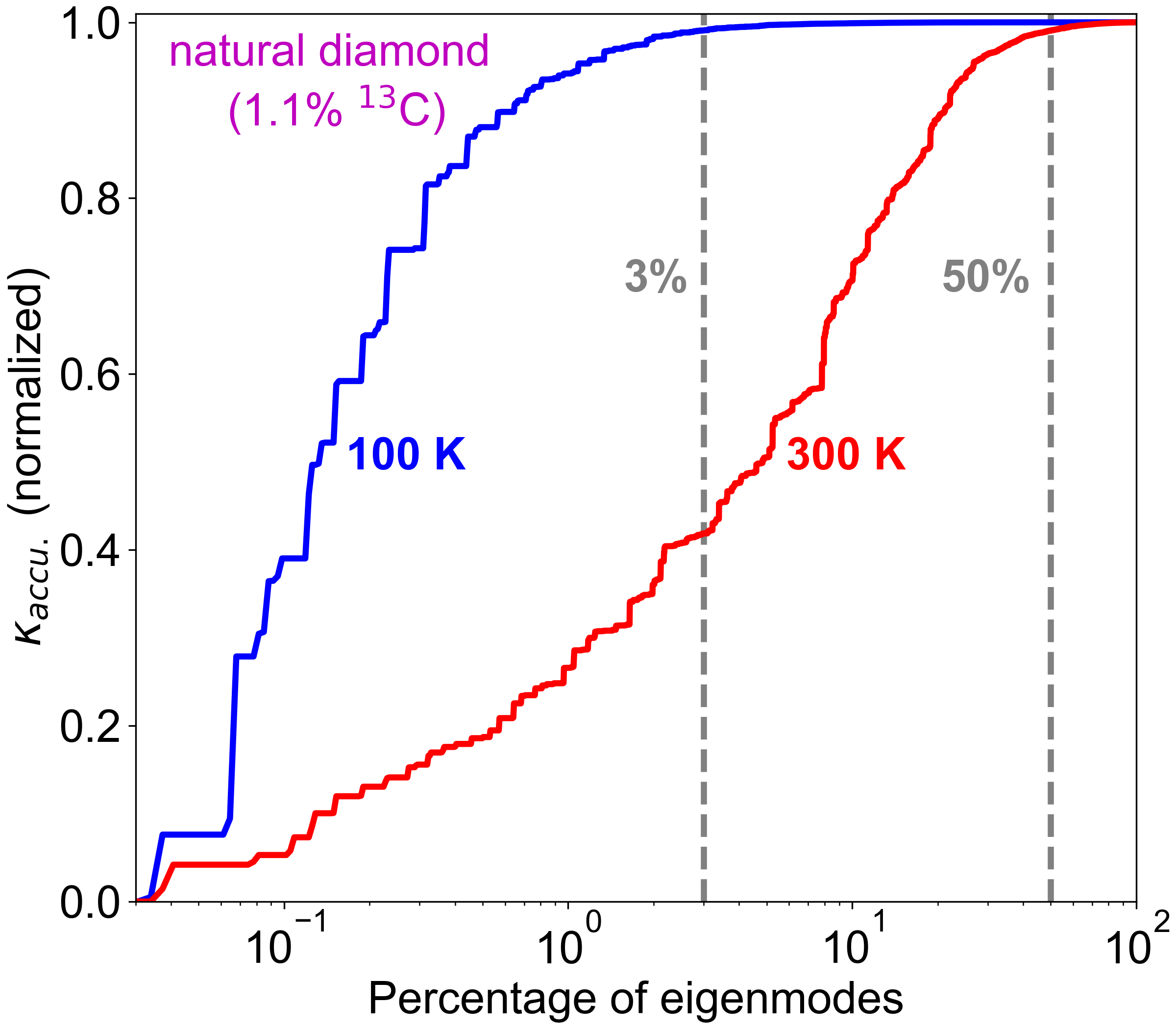}
    \caption{Thermal conductivity accumulation ($\kappa_{accu.}$) vs. percentage of eigenmodes, arranged in the increasing order of their eigenvalues, for natural diamond at 300 K [red] and 100 K, [blue]. While 50\% of the eigenmodes of $\Omega$ contribute to the $\kappa$ of naturally-occurring diamond at 300 K, only about 3\% of them contribute to more than 99\% of the $\kappa$ at 100 K, thus facilitating a low-rank description of the steady-state thermal transport at low temperatures.}
    \label{fig:kappa_accu}
\end{figure}

Figure~\ref{fig:kappa_accu} shows the normalized cumulative contribution to $\kappa$ from the eigenmodes of $\Omega$ for diamond in its naturally occurring form, calculated from Eq.~\ref{eq:eigenmode_kappa}. In this figure, the eigenmodes are sorted in the increasing order of their respective eigenvalues. At 300 K, the contributions from just half of the eigenmodes accumulate to 99\% of the total $\kappa$, indicating that the rest of the spectrum of $\Omega$ is predominantly composed of numerically even eigenmodes. This fraction of numerically even eigenmodes increases to more than 97\% at 100 K, since the contributions to 99\% of the total $\kappa$ come from only $\sim 3\%$ of the eigenmodes of $\Omega$ (Fig.~\ref{fig:kappa_accu}). Additionally, the eigenmodes of $\Omega$ with large eigenvalues have their contributions to $\kappa$ suppressed by the appearance of the eigenvalues in the denominator in Eq.~\ref{eq:eigenmode_kappa}. This feature of the spectrum of $\Omega$ naturally motivates a low-rank representation of the LPBE system, where the contributions from a small fraction of the eigenmodes with large $\mathcal{V}^{0m}/\sqrt{\sigma^m}$ are sufficient to accurately capture $\kappa$. It is worth noting here that we cannot achieve such a reduction in the dimensionality of the LPBE system in the conventional phonon picture, since all phonons are intricately coupled by phonon-phonon scattering events.\\

This rank reduction naturally extends to the transient LPBE system as well, due to the similarity in the explicit dependence of the two solutions (Eqs.~\ref{eq:eigenmode_kappa} and~\ref{eq:f_delta_T}) on the velocities and the eigenvalues of the eigenmodes of $\Omega$. The additional complexity of the dependence of $\mathcal{P}^{mn}\left(\xi, \eta\right)$ in Eq.~\ref{eq:f_delta_T} on the even eigenmodes of $\Omega$ is also alleviated by noting that only those eigenmodes of $\Omega$ with small eigenvalues contribute significantly to the intermediate matrix $\Psi$, and so, to $\mathcal{P}^{mn}\left(\xi, \eta\right)$, as detailed in the Appendix~\ref{sec:low_rank_lpbe}. Hence, it is sufficient to consider a rank-reduced intermediate matrix $\Psi_R$, which is a much smaller block carved out of $\Psi$, contributed only by a few eigenmodes of $\Omega$ with the smallest eigenvalues. Noting that $\Psi$ must be diagonalized repeatedly for different values of $\xi$ and $\eta$ to obtain the temporal temperature response, $\Delta\tilde{T}\left(\xi, \eta\right)$, the reduction in dimension from $\Psi$ to $\Psi_{R}$ results in a dramatic reduction in the computational cost, since the cost of diagonalization scales as the cubic power of the size of the matrix.\\

\section{Accelerating prediction of transient phonon dynamics using the low-rank representation of the LPBE}
We demonstrate the acceleration of the LPBE solution achieved through the low-rank representation by predicting the temporal temperature dynamics from the transient grating (TG) experiments on diamond at 100 K. Apart from being the experiment of choice for several recent observations of the second sound~\cite{huberman_observation_2019, ding_observation_2022}, TG measurements can be obtained for different spatial wave-vectors ($\xi$) independently, thus enabling direct experimental comparisons with our predicted temporal temperature dynamics in the future. Furthermore, the ability of the TG to explicitly set and vary the spatial wave-vector $\xi$ in a non-contact manner facilitates the realization of diffusive, hydrodynamic and ballistic heat flow regimes, all in the same sample, as described in several previous works~\cite{johnson_direct_2013, ravichandran_spectrally_2018, huberman_observation_2019, ding_observation_2022}. For our study, we choose diamond - the material with the highest $\kappa$ known to date, and perform calculations for naturally occurring and isotopically enriched crystals at 100 K, where the existing measurements of $\kappa$ in the literature suggest dominant intrinsic phonon-phonon scattering due to crystal anharmonicity, and show negligible effects of extrinsic phonon scattering due to defects and boundaries in the samples~\cite{berman_nitrogen_1975, onn_aspects_1992, olson_thermal_1993, wei_thermal_1993}. \\

\begin{figure}[h]
    \centering
    \includegraphics[width=1.0\linewidth]{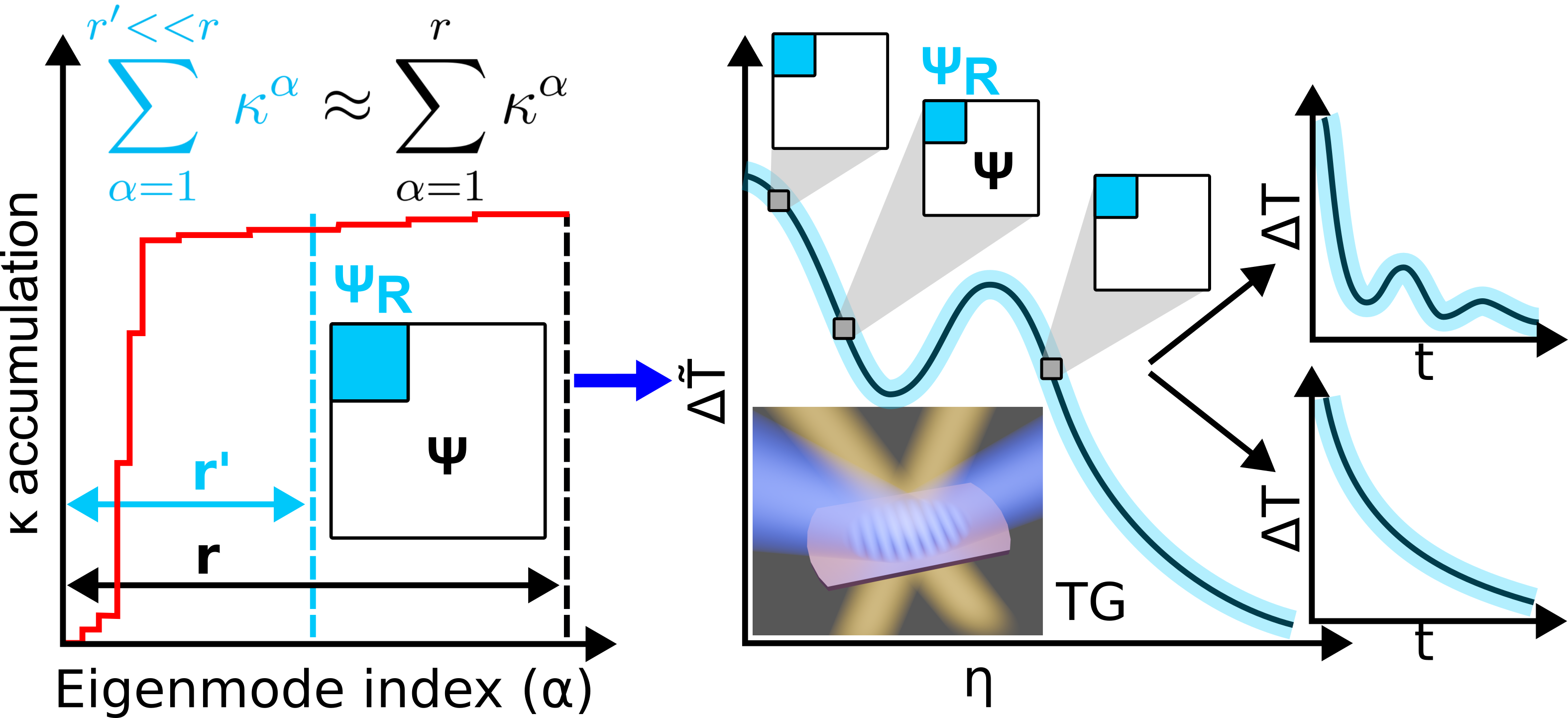}
    \caption{Workflow for predicting the TG temperature dynamics using the low-rank representation of the LPBE. First, the collision matrix $\Omega$ is diagonalized, and the number of eigenmodes of $\Omega$ with the smallest eigenvalues, required to achieve convergence on the steady-state $\kappa$, is obtained from the $\kappa$-accumulation curve. Next, the LPBE is solved in the frequency domain for each temporal frequency ($\eta$) and spatial grating wave-vector ($\xi$) with number of eigenmodes of the collision matrix selected from the $\kappa$-accumulation curve. The temporal temperature responses for different $\xi$’s are obtained from the inverse Fourier transforms of $\Delta \tilde{T}\left(\xi, \eta\right)$ over the variable $\eta$. In the plots of the frequency-domain and time-domain temperature responses, the thin black curves represent the full-rank solutions of LPBE obtained by diagonalizing the full intermediate matrix $\Psi$ for each value of $\eta$, and the corresponding low-rank solutions, obtained by diagonalizing the rank-reduced intermediate matrix $\Psi_R$, are shown by the thick blue curves.}
    \label{fig:low_rank_illustration}
\end{figure}

Figure~\ref{fig:low_rank_illustration} illustrates a summarizing work-flow for our low-rank solution of the LPBE. First, we diagonalize the collision matrix $\Omega$ and sort the eigenmodes in the increasing order of their eigenvalues. Next, we construct an accumulation curve for $\kappa$ as a function of the eigenvalues, and select those eigenmodes whose contributions accumulate to within $1\%$ of the total $\kappa$. Then, we employ these selected eigenmodes to construct the low-rank intermediate matrix $\Psi_{R}$ for each $\{ \mathbf{\xi}, \eta \}$ pair, and obtain the Fourier-transformed temperature deviation $\Delta \tilde{T} \left( \mathbf{\xi}, \eta \right)$ [Eq.~\ref{eq:f_delta_T}]. As discussed below, for most of the cases, only a subset of these selected eigenmodes are necessary to construct $\Psi_R$ to achieve optimal convergence on the transient temperature dynamics. Finally, we perform an inverse Fourier transform over the variables $\eta$ and $\xi$ to obtain the spatially-resolved transient temperature response - $\Delta T \left( \textbf{x}, t \right)$.\\

\begin{figure}[h]
    \centering
    \includegraphics[width=1.0\linewidth]{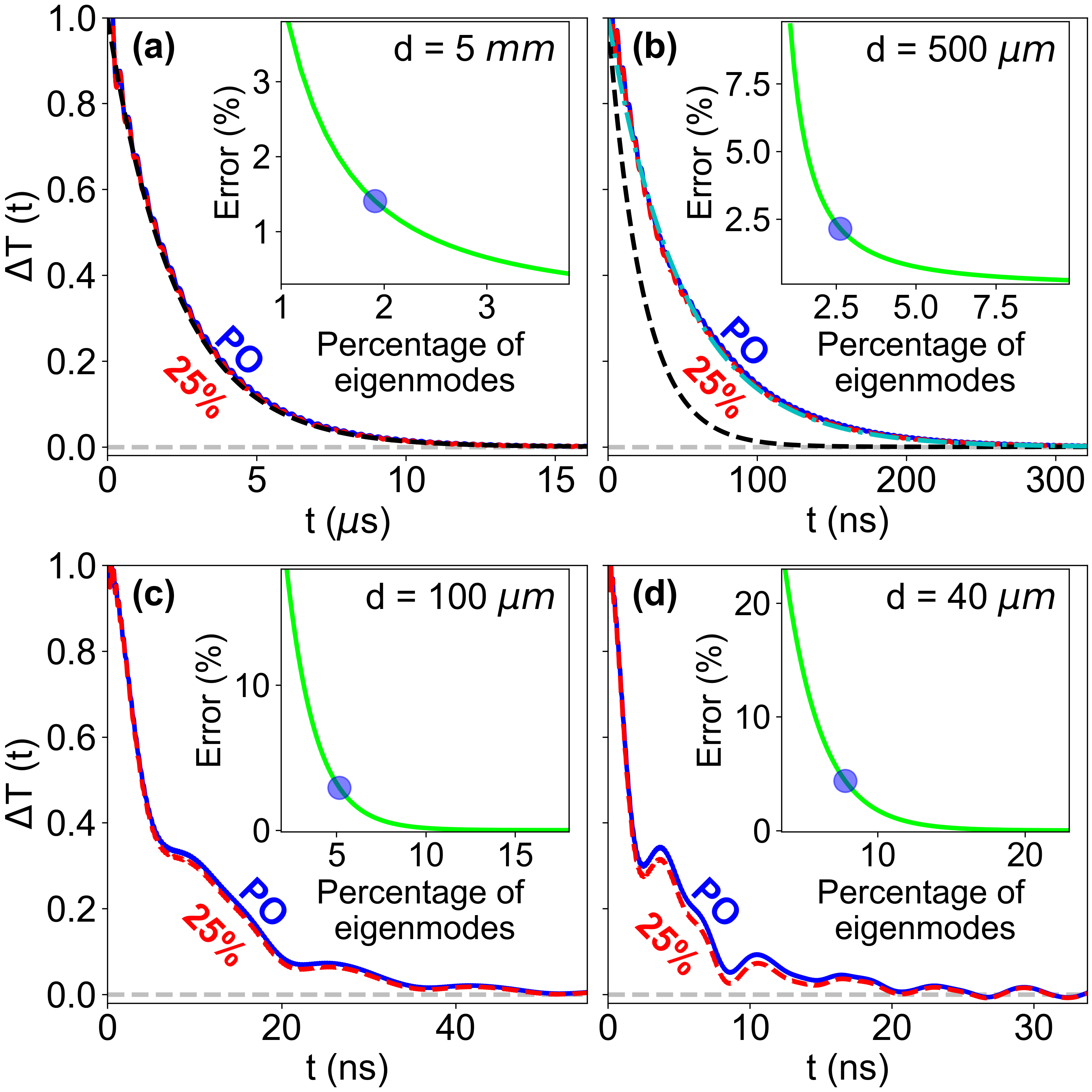}
    \caption{Simulated temperature responses to impulsive heating in the TG experiment on naturally occurring diamond at 100 K for four different heat flow regimes. The black-dashed curves in panels~(a) and (b) represent the solution of the classical Fourier heat equation with the bulk $\kappa$, and the cyan dashed-dotted curve in (b) is a fit to the Fourier heat equation with an effective lower $\kappa$. The Fourier-diffusive (a) and the quasiballistic (b) regimes fit well with the solution of the classical Fourier heat equation with the bulk $\kappa$ and an effective lower $\kappa$, respectively. The oscillatory features in panels~(c) and (d) are signatures of weak hydrodynamic and ballistic heat flow regimes respectively. The optimal number of eigenmodes required to achieve convergence on the transient temperature dynamics is determined from the pareto-optimality condition, shown as a blue dot in the inset for each regime. The gold-standard LPBE solutions against which the errors in the insets are estimated, are shown by the red-dashed curves and are computed by including 25\% of the eigenmodes of $\Omega$ - far more than what is needed to achieve convergence on the $\kappa$ and the transient temperature response. }
    \label{fig:TG_simulation}
\end{figure}

Figure~\ref{fig:TG_simulation} shows the transient temperature response to impulsive heating of natural diamond (1.1\% $^{13}C$) in TG at 100 K over a range of grating periods ($d$) from 40 $\mu$m till 5 mm. Here, the red-dashed lines represent the solution of LPBE considering 25\% of the total number of eigenmodes of $\Omega$ - significantly more than that needed to achieve convergence of $\kappa$ in Fig.~\ref{fig:kappa_accu}, and the blue lines represent the solution with the optimal number of eigenmodes, which are estimated from the pareto-optimality (PO) condition as shown in the inset of each panel. The PO point signifies an optimal trade-off between the effort - the percentage of the total number of eigenmodes of $\Omega$ used in the solution of the LPBE, and the reward - the error in the transient temperature profile relative to that computed with $25\%$ of the total number of eigenmodes. We find that the PO condition is met with less than 7\% of the total number of eigenmodes of $\Omega$ in all four cases shown in Fig.~\ref{fig:TG_simulation} (see Supplementary section~S4 for detailed convergence plots).\\

\begin{figure}[h]
    \centering
    \includegraphics[width=1.0\linewidth]{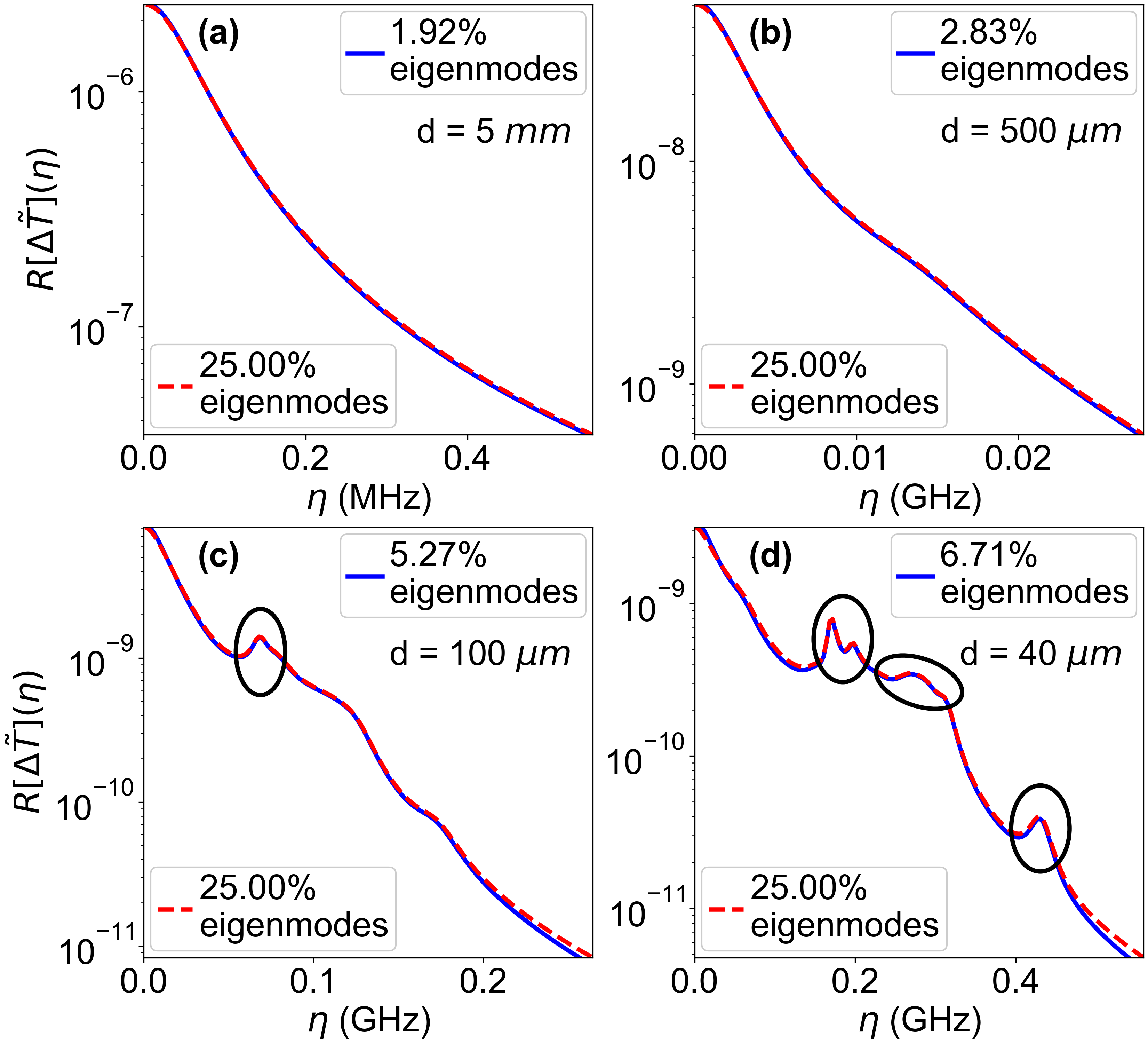}
    \caption{Fourier transforms of the simulated temperature response to impulsive heating in the TG experiment for the four different heat flow regimes shown in Fig.~\ref{fig:TG_simulation}. The frequency-domain solution of the LPBE for the Fourier-diffusive (a) and the quasiballistic (b) regimes are Lorentzians centered at zero frequency (only the positive frequencies are shown in all four panels for clarity). A single small peak superimposed over a background Lorentzian in panel~(c) is a signature of the weak hydrodynamic regime while multiple peaks in panel~(d) correspond to the ballistic heat flow regime. The blue curves in all four panels represent the pareto-optimal solutions, while the red-dashed curves are obtained by considering 25\% of the eigenmodes of $\Omega$, as in Fig.~\ref{fig:TG_simulation}.}
    \label{fig:TG_simulation_fourier_domain}
\end{figure}

In Fig.~\ref{fig:TG_simulation}(a), for a TG period of 5 mm, the solution of the LPBE fits well to the solution of the Fourier's heat diffusion equation for the TG experiment, which is shown by the black-dashed line - an exponential decay in time with a decay rate given by $4 \pi^{2} \alpha / d^{2}$, where $\alpha$ is the thermal diffusivity of the material. When $d$ is reduced to 500 $\mu$m, the quasiballistic heat transfer regime is observed, where the LPBE solution deviates from the Fourier's solution of heat diffusion; however, it is still an exponential decay with a decay rate smaller than the prediction from the Fourier's heat diffusion equation, as shown by the cyan dashed-dotted curve in Fig.~\ref{fig:TG_simulation}(b)~\cite{hua_transport_2014, ravichandran_role_2016}. For these two regimes, the frequency-domain solutions are Lorentzians centered at zero frequency as shown in Figs.~\ref{fig:TG_simulation_fourier_domain}(a) and (b) respectively.\\

Besides these two regimes, our calculations also uncover two different types of damped oscillatory temperature decay for $d$=100 $\mu m$ (Fig.~\ref{fig:TG_simulation}(c)) and 40 $\mu m$ (Fig.~\ref{fig:TG_simulation}(d)). For the former case, the corresponding frequency-domain solution in Fig.~\ref{fig:TG_simulation_fourier_domain} (c) shows a single prominent broad peak at a temporal frequency of about 55 MHz, indicative of collective hydrodynamic phonon transport with a single drift velocity~\cite{huberman_observation_2019}. In the latter case, where the TG period is reduced further to 40 $\mu m$, we find multiple narrow peaks in the frequency-domain solution of the LPBE, as shown in Fig.~\ref{fig:TG_simulation_fourier_domain}(d), indicating the onset of scatter-free ballistic transport of phonons across the TG period, with the different peaks corresponding to different phonon modes with different magnitudes of group velocities along the direction of the temperature gradient. Thus, our low-rank solution of the LPBE is able to predict the conditions for diffusive, quasiballistic, hydrodynamic and ballistic phonon transport regimes without any ad-hoc adjustments to the computational scheme.\\

\section{Uncovering amplification of phonon hydrodynamics upon isotopic enrichment}
\begin{figure}
    \centering
    \includegraphics[width=1.0\linewidth]{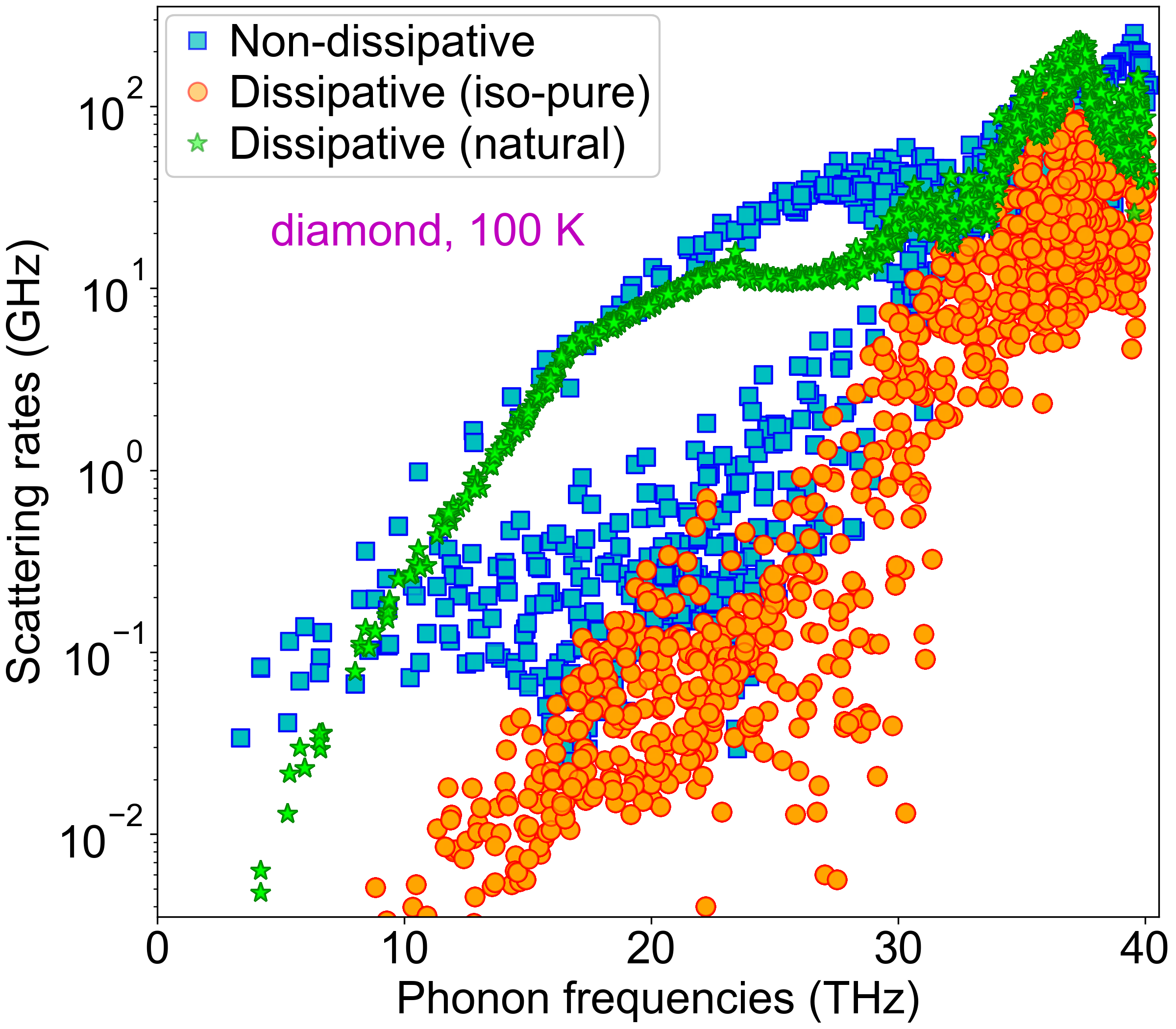}
    \caption{Scattering rates for the total dissipative and non-dissipative (anharmonic-Normal) processes in diamond at 100 K. For pure diamond, the only dissipative processes are the anharmonic-Umklapp scattering events, but for natural diamond, the dissipative processes include the phonon-isotope scattering events as well. For natural diamond, the dissipative scattering rates (green stars) are comparable to or larger than those of the non-dissipative events (blue squares). Thus, the second sound signatures are weaker in natural diamond [Fig.~\ref{fig:TG_simulation} (c)]. When the phonon-isotope scattering events are excluded for the isotopically pure case, the resulting dissipative scattering rates (orange circles) are significantly weaker than their non-dissipative counterparts; hence, a strong amplification of the second sound signature can be expected upon isotopic enrichment of diamond at 100 K.}
    \label{fig:scattering_rates}
\end{figure}

Focusing on the hydrodynamic signatures seen in Fig.~\ref{fig:TG_simulation} (c), we notice that the oscillations in the TG temperature response for naturally-occurring diamond are rather weak and may not be discernible in the presence of experimental noise. This suppressed hydrodynamic signature is caused by strongly momentum-dissipative scattering of phonons by the isotopic mixture of carbon atoms in naturally-occurring diamond, even though anharmonic scattering is dominated by the non-dissipative Normal scattering processes, as shown in Fig.~\ref{fig:scattering_rates}. Since many of the ultrahigh-$\kappa$ materials have been reported to have strong isotope effect on their $\kappa$~\cite{lindsay_thermal_2012, zheng_high_2018, chen_ultrahigh_2020}, our results suggest that isotopic enrichment should amplify the hydrodynamic signatures in these materials, even at 100 K.\\

To test our hypothesis, we applied the low-rank solution of the LPBE to diamond with the concentration of the $^{12}$C isotope enriched to 99.95\% - the highest isotopic purity of diamond on which the measurements of $\kappa$ have been reported in the literature~\cite{onn_aspects_1992}. Unlike the natural-occurring case, the steady-state $\kappa$ of isotopically-enriched diamond at 100 K achieves convergence to within $\sim$ 10\% for a phonon Brillouin zone discretization of 35$^3$ (as shown in the Supplementary Fig.~S1), thus resulting in a collision matrix $\Omega$ of size $\sim$ 0.3 million $\times$ 0.3 million. This requirement of a refined Brillouin zone density originates from the significantly suppressed momentum-dissipating phonon-isotope scattering events upon isotopic enrichment (as shown in the Supplementary Fig.~S3). Complete diagonalization of such a large, dense collision matrix is prohibitively expensive, with the estimates of diagonalization time exceeding $\sim 48$ months of CPU time on the state-of-the-art computing hardware (see Supplementary section~S3 for details about the dimension of $\Omega$ and the computational challenges). However, since we require only a small fraction of the eigenmodes of $\Omega$ with the smallest eigenvalues to obtain the steady-state $\kappa$ and solve the transient LPBE, we find it far more efficient to use Krylov methods such as the implicitly-restarted Arnoldi method (IRAM)~\cite{lehoucq_deflation_1996}, which can compute the first few eigenmodes of a matrix with the smallest eigenvalues at a fraction of the computational cost of the conventional full-scale diagonalization algorithms. Additionally, we leverage the symmetries of $\Omega$ derived from the space-group symmetry of the diamond structure to reduce the requirement on the computational memory from $\sim$ 1.0 terabyte (TB) for the conventional diagonalization to about 0.02 TB for our low-rank LPBE solution, and to take advantage of the matrix-free Krylov algorithms for diagonalization (see Appendix~\ref{sec:matrix_free_approch} for details on the matrix-free diagonalization protocol).\\

% \begin{figure}
%     \centering
%     \includegraphics[width=1.0\linewidth]{./images/enriched_diamond.png}
%     \caption{Simulated temperature response to impulsive heating in TG for $d$=200 $\mu m$ at 100 K in isotopically-enriched diamond (with 0.05\% $^{13}C$) showing the hydrodynamic second sound regime. To obtain this result, we used $\sim$ 0.5\% of the eigenmodes of the collision matrix, which was sufficient to obtain convergence to within 1\% of the steady-state $\kappa$, as shown by the blue curve in the inset. The 100\% converged $\kappa$ for the isotopically-enriched diamond is obtained from the conventional phonon picture as described in ref.~\cite{malviya_failure_2023}.}
%     \label{fig:enriched_diamond}
% \end{figure}

\begin{figure}
    \centering
    \includegraphics[width=1.0\linewidth]{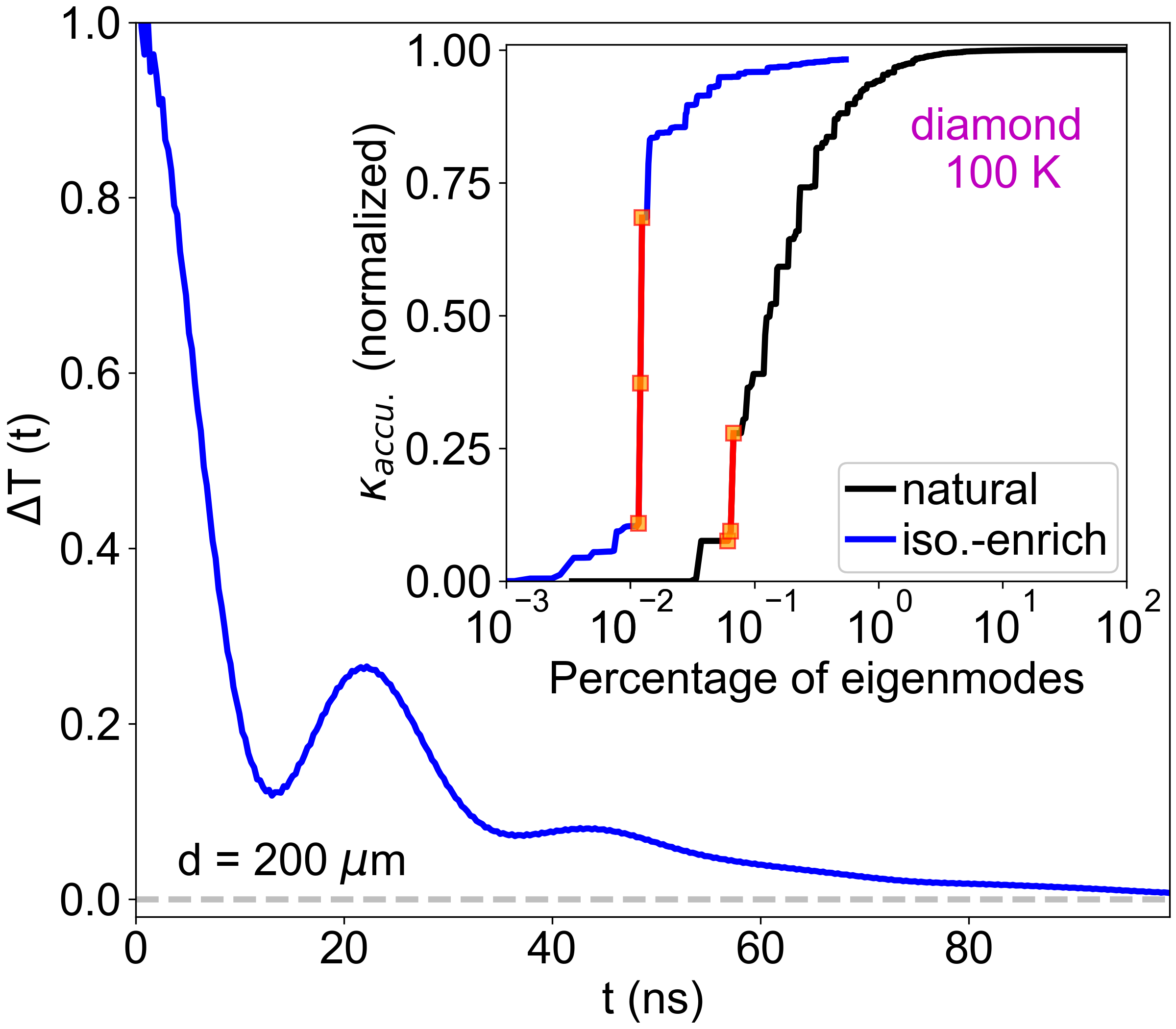}
    \caption{Simulated temperature response to impulsive heating in TG for $d$=200 $\mu m$ at 100 K in isotopically-enriched diamond (with 0.05\% $^{13}C$) showing the hydrodynamic second sound regime. To obtain this result, we used $\sim$ 0.5\% of the eigenmodes of the collision matrix, which was sufficient to obtain convergence to within 1\% of the steady-state $\kappa$, as shown by the blue curve in the inset. The 100\% converged $\kappa$ for the isotopically-enriched diamond is obtained from the conventional phonon picture as described in ref.~\cite{malviya_failure_2023}.}
    \label{fig:enriched_diamond}
\end{figure}

We employ the matrix-free IRAM algorithm to obtain 1400 eigenmodes of $\Omega$ with the smallest eigenvalues for isotopically enriched diamond at 100 K, which corresponds to $\sim$ 0.5\% of its entire spectrum. We find that upon isotopic enrichment, an even smaller fraction of the eigenmodes of $\Omega$ are sufficient to achieve convergence on $\kappa$, compared to the naturally-occurring case. As shown in the inset of Fig.~\ref{fig:enriched_diamond}, while 3\% of the eigenmodes of $\Omega$ contribute to $\sim$ 99\% of the total $\kappa$ of natural diamond, just 0.5\% of the eigenmodes are needed for the isotopically enriched case to achieve the same level of convergence. Additionally, we find that, upon isotopic enrichment, the fractional contribution to $\kappa$ from three dominant eigenmodes, shown by the red segments in the accumulation curves in the inset of Fig.~\ref{fig:enriched_diamond}, increases from 20\% for the natural case to about 70\% for the isotopically enriched case. This transition is consistent with the limiting case of purely non-dissipative Normal scattering, where the three drifting eigenmodes of the collision matrix completely dominate the contribution to $\kappa$~\cite{guyer_solution_1966, hardy_phonon_1970, pitaevskii_physical_2012}.\\

Using these 1400 eigenmodes of $\Omega$ with the smallest eigenvalues for the isotopically enriched diamond at 100 K, we simulate the TG temperature response, and observe strong amplification of the second sound oscillations compared to those observed for the naturally-occurring material, as can be seen from Fig.~\ref{fig:enriched_diamond} for a TG period of 200 $\mu$m. Interestingly, we find that the oscillatory signature directly originates from the three dominant eigenmodes of $\Omega$ highlighted in the inset of Fig.~\ref{fig:enriched_diamond}. When we artificially remove these eigenmodes from the calculation, we find that the oscillations are suppressed, as shown in the Supplementary Fig.~S4. Our results suggest that the dominance of three eigenmodes of $\Omega$ to $\kappa$ can be used as a strong indicator for the possibility of hydrodynamic thermal transport in cubic materials, even when the momentum-dissipating phonon scattering processes are weak, but not necessarily vanishingly small. \\

\section{Conclusions and discussion} \label{sec:discussion}
In summary, we discover a low-rank representation of the linearized Peierls-Boltzmann equation for phonon transport, originating from the spectral properties of the phonon collision matrix, that enables rapid first-principles prediction of the ultrafast transient temperature dynamics in electrical insulators at a fraction of the computational cost of the conventional brute-force methods, without compromising on the accuracy. In particular, our computational approach allows for a fully first-principles temporal tracking of strongly-coupled phonon systems with the number of degrees of freedom exceeding quarter of a million, which would otherwise be intractable using state-of-the-art computational techniques. By studying the transient temperature dynamics in diamond at 100 K, we highlight that such stringent computational requirements often occur in common materials at experimentally accessible conditions, and predict the conditions to realize Fourier-diffusive as well as novel nondiffusive phonon transport regimes, all within the same material. We leverage the symmetry of the phonon collision matrix derived from the space-group operations of the underlying crystal to further accelerate the computations for large degree-of-freedom systems by employing the matrix-free Krylov-subspace diagonalization algorithms.\\

Our accelerated first-principles framework has enabled the identification of the key role played by isotopic enrichment in amplifying hydrodynamic second sound signatures in the predicted transient temperature dynamics, through the example of diamond - the material with the highest $\kappa$ known till date. Using this framework, we obtained simple guidelines, drawn from the contribution of the dominant eigenmodes of the phonon collision operator to $\kappa$, to predict the possibility of and the necessary conditions for hydrodynamic phonon transport in common semiconductors.\\ 

Our low-rank solution is centered on the linearity of the underlying dynamical system that enables a spectral decomposition of the governing equations, and is particularly effective for large degree-of-freedom systems. These features are often observed in the coupled dynamics of quasiparticles in condensed matter with vastly different energy scales, whose transport is governed by the semi-classical coupled Boltzmann equations. For example, coupled dynamics of electrons and phonons, that underpins novel energy transport processes such as phonon drag, ultrafast electron thermalization and the phonon bottleneck effect, often require ultrafine discretization of the phonon and electron Brillouin zones, to resolve the energy-conserving collisions between electrons with energies in eV and phonons with energies in meV. Our low-rank approach will enable computationally-efficient fully first-principles solutions for such seemingly intractable problems that will drive future materials discovery efforts.

\section*{Author contributions}
N.K.R. originated the research idea. N.M. developed the computational framework and performed the calculations. N.M. and N.K.R. analyzed the results and wrote the manuscript.

\begin{acknowledgments}
This work was supported by the Core Research Grant (CRG) No. CRG/2020/006166, and the Mathematical Research Impact Centric Support (MATRICS) grant no. MTR/2022/001043 from the Department of Science and Technology - Science and Engineering Research Board, India. NM gratefully acknowledges the Prime Minister's Research Fellowship (PMRF) grant no. PMRF-02-01036. NR thanks the Infosys Foundation for their support through a Young Investigator Award.
\end{acknowledgments}

\appendix
\section{First-principles computations of the phonon collision matrix} \label{sec:phonon_collision_matrix}
In our calculations, we have incorporated anharmonic three-phonon and phonon-isotope scatterings to construct the collision matrix. The linearized three-phonon collision matrix ($\mathcal{R}^\text{3ph.}$), is given as~\cite{broido_lattice_2005, ravichandran_phonon-phonon_2020, malviya_failure_2023}:
\begin{align}
    \mathcal{R}_{\lambda \lambda_{1}}^\text{3ph.} = &
    \left[ \sum_{\lambda_{1} \lambda_{2}} \left( \mathcal{W}_{\lambda \lambda_{1} \lambda_{2}}^{+} + \frac{1}{2} \mathcal{W}_{\lambda \lambda_{1} \lambda_{2}}^{-} \right) \right] \Delta_{\lambda \lambda_{1}} \nonumber \\
    & + \left[ \sum_{\lambda_{2}} \left( \mathcal{W}_{\lambda \lambda_{1} \lambda_{2}}^{+} - \mathcal{W}_{\lambda \lambda_{1} \lambda_{2}}^{-} - \mathcal{W}_{\lambda_{1} \lambda \lambda_{2}}^{-} \right) \right]
    \label{eq:lin_3ph_collision_matrix}
\end{align}
where $\mathcal{W}_{\lambda \lambda_{1} \lambda_{2}}^{\pm}$ are the transition rates for phonon emission $\left( - \right)$ and absorption $\left( + \right)$ given by: 
\begin{align}
    \label{eq:pm_transition_rates}
    \mathcal{W}_{\lambda \lambda_{1} \lambda_{2}}^{\pm} 
    = & \frac{\pi \hbar}{4N_{0}} 
    \frac{ \vert \mathbf{\mathcal{X}}_{\lambda \left( \pm \lambda_{1} \right) \left(-\lambda_{2}\right)} \vert^{2}}{\omega_{\lambda} \omega_{\lambda_{1}} \omega_{\lambda_{2}}} 
    \delta\left( \omega_{\lambda} \pm \omega_{\lambda_{1}} - \omega_{\lambda_{2}}\right) \nonumber \\
    & \times f_{\lambda}^{0} \left( f_{\lambda_{1}}^{0} + \frac{1}{2} \mp \frac{1}{2} \right) \left( f_{\lambda_{2}}^{0} + 1 \right)
\end{align}
and $\Delta_{\lambda\lambda_1}$ is the Kronecker delta function. The $\mathcal{X}_{\lambda \lambda_{1} \lambda_{2}}$'s are related to the third order (cubic) inter-atomic force constants (IFCs) $\Phi_{\alpha \beta \gamma} \left( l k, l_{1} k_{1}, l_{2} k_{2} \right)$ as:
\begin{align}
    \mathcal{X}_{\lambda \lambda_{1} \lambda_{2}} 
    = &
    \frac{1}{N_{0}} \sum_{l l_{1} l_{2}} \sum_{k k_{1} k_{2}} \sum_{\alpha \beta \gamma} 
    \frac{\Phi_{\alpha \beta \gamma} \left( l k, l_{1} k_{1}, l_{2} k_{2} \right)}{\sqrt{m_{k} m_{k_{1}} m_{k_{2}}}} \nonumber \\
    & \times w_{\alpha}\left( \textbf{q}j, k \right) w_{\beta}\left( \textbf{q}_{1} j_{1}, k_{1} \right) w_{\gamma}\left( \textbf{q}_{2} j_{2}, k_{2} \right) \nonumber \\
    & \times \exp{\left( i \textbf{q} \cdot \textbf{R}\left( l \right) \right)}
    \exp{\left( i \textbf{q}_{1} \cdot \textbf{R}\left( l_{1} \right) \right)} \nonumber \\
    & \times \exp{\left( i \textbf{q}_{2} \cdot \textbf{R}\left( l_{2} \right) \right)}
    \label{eq:fourier_transformed_force_constant}
\end{align}
where $N_{0}$ is the number of unit cells in the crystal, $\alpha$, $\beta$, $\gamma$ are the Cartesian directions, $k$, $k_1$, $k_2$ are the basis atoms with masses $m_{k}$, $m_{k_1}$, $m_{k_2}$ in the $l$, $l_1$, $l_2$ unit cells at equilibrium positions $\textbf{R} \left( lk \right)$, $\textbf{R} \left( l_1k_1 \right)$, $\textbf{R} \left( l_2k_2 \right)$, respectively, and $w$ is the eigenvector of the dynamical matrix. We have not included higher-order scattering among four phonons in this work, as it has been shown to have a negligible effect on the $\kappa$ of diamond at low temperatures~\cite{ravichandran_unified_2018}. \\

We have used the first-principles approach based on density functional theory (DFT) as described in ref.~\cite{ravichandran_unified_2018} to obtain the harmonic and cubic interatomic force constants required to compute $\mathcal{W}_{\lambda \lambda_{1} \lambda_{2}}^{\pm}$. The converged parameters for these DFT calculations are enlisted in ref.~\cite{malviya_failure_2023}. \\

The collision matrix $\mathcal{R}^{\text{ph.-iso.}}$~\cite{tamura_isotope_1983} for phonon-isotope scattering is given by:
\begin{equation}
    \mathcal{R}_{\lambda \lambda_{1}}^{\text{ph.-iso.}} = \sum_{\lambda_{1}} \mathcal{W}_{\lambda \lambda_{1}}^{\text{ph.-iso.}} \Delta_{\lambda \lambda_{1}} - \mathcal{W}_{\lambda \lambda_{1}}^{\text{ph.-iso.}}
    \label{eq:lin_ph-iso_collision_matrix}
\end{equation}
where the phonon-isotope transition rates $\mathcal{W}_{\lambda \lambda_{1}}^{\text{ph.-iso.}}$ are given by:
\begin{align}
    \mathcal{W}_{\lambda \lambda_{1}}^{\text{ph.-iso.}} = & \frac{\pi \omega_{\lambda}^{2} f_{\lambda}^0 \left( f_{\lambda}^0+1 \right)}{2 N_{0}} \vert \Phi_{\lambda \left(- \lambda_{1} \right)}^{\text{ph.-iso.}} \vert ^2 \delta \left( \omega_{\lambda}-\omega_{\lambda_{1}} \right)
    \label{eq:ph-iso_transition_rates}
\end{align}
with the matrix elements $\vert \Phi_{\lambda \left(- \lambda_{1} \right)}^{\text{ph.-iso.}} \vert ^2$ given by:
\begin{gather}
    \vert \Phi_{\lambda \left(- \lambda_{1} \right)}^{\text{ph.-iso.}} \vert^{2} 
    = \sum_{k} \mathfrak{g} \left( k \right) \vert w\left( k \vert \textbf{q} j \right) \cdot w^{*}\left( k \vert \textbf{q}' j'\right) \vert^{2}
\end{gather}
and the mass variance parameter, $\mathfrak{g} \left( k \right)$, is given by:
\begin{gather}
    \mathfrak{g} \left( k \right) 
    = \sum_{i}  \mathfrak{f}_{i} \left( k \right) \left(1 - \frac{m_{k,i}}{\overline{m}_{k}} \right) ^{2}
\end{gather}
with $m_{k,i}$ being the mass of the $i^{\mathrm{th}}$ isotope of the $k^{\mathrm{th}}$ atom, having a fractional concentration of $\mathfrak{f}_{i} \left( k \right)$, and $\overline{m}_{k}$ being the corresponding average mass considering the isotopic mixture.\\

Finally, we have used the following transformation to get the collision matrix $\Omega$, which is used in the main text, from the Eqs.~\ref{eq:lin_3ph_collision_matrix} and~\ref{eq:lin_ph-iso_collision_matrix}:
\begin{equation}
    \Omega_{\lambda \lambda_{1}} = \frac{\mathcal{R}_{\lambda \lambda_{1}}^\text{3ph.} + \mathcal{R}_{\lambda \lambda_{1}}^{\text{ph.-iso.}}}{\displaystyle \sqrt{f_{\lambda}^{0} \left( f_{\lambda}^{0} + 1 \right)} \sqrt{f_{\lambda_{1}}^{0} \left( f_{\lambda_{1}}^{0} + 1 \right)}}
\end{equation}

\section{Low-rank solution of the LPBE} \label{sec:low_rank_lpbe}
\begin{figure}[h]
    \centering
    \includegraphics[width=1.0\linewidth]{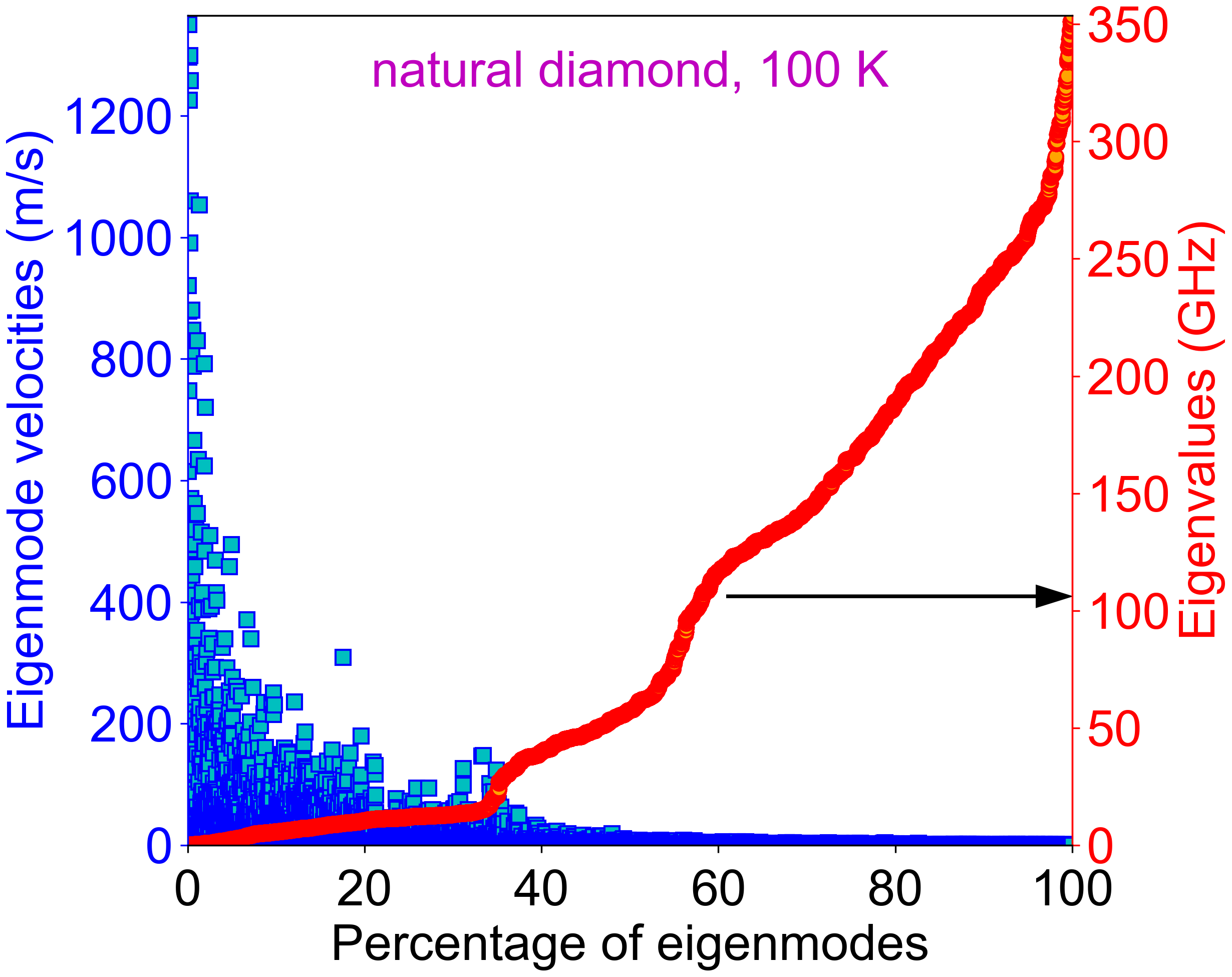}
    \caption{The magnitude of the velocities and the eigenvalues of the eigenmodes of $\Omega$ for natural diamond at 100 K. The velocities of the eigenmodes drop sharply as the eigenvalues increase, thus resulting in a weaker contribution to the steady-state $\kappa$ and the transient $\Delta T$ from the eigenmodes with large eigenvalues, thereby motivating a low-rank representation of the LPBE system.}
    \label{fig:velocity_eigenvalue}
\end{figure}

As discussed in section~\ref{sec:results}, for naturally occurring diamond, approximately $3 \%$ of the eigenmodes of $\Omega$ contribute to 99\% of $\kappa$ at 100 K (see Fig.~\ref{fig:kappa_accu}), thereby motivating a low-rank representation of the LPBE system. This sharp contribution to $\kappa$ from a small fraction of the eigenmodes with the smallest eigenvalues is caused by the precipitous drop in the velocities of the eigenmodes with increasing eigenvalues, as shown in Fig.~\ref{fig:velocity_eigenvalue}. Thus, the term $\mathcal{V}^{0m}/\sqrt{\sigma^m}$, through which the eigenmodes of $\Omega$ contribute to $\kappa$, diminishes rapidly with increasing eigenvalues, thus shifting the dominant contribution to $\kappa$ to the lower end of the spectrum of $\Omega$.\\

\begin{figure}[h]
    \centering
    \includegraphics[width=1.0\linewidth]{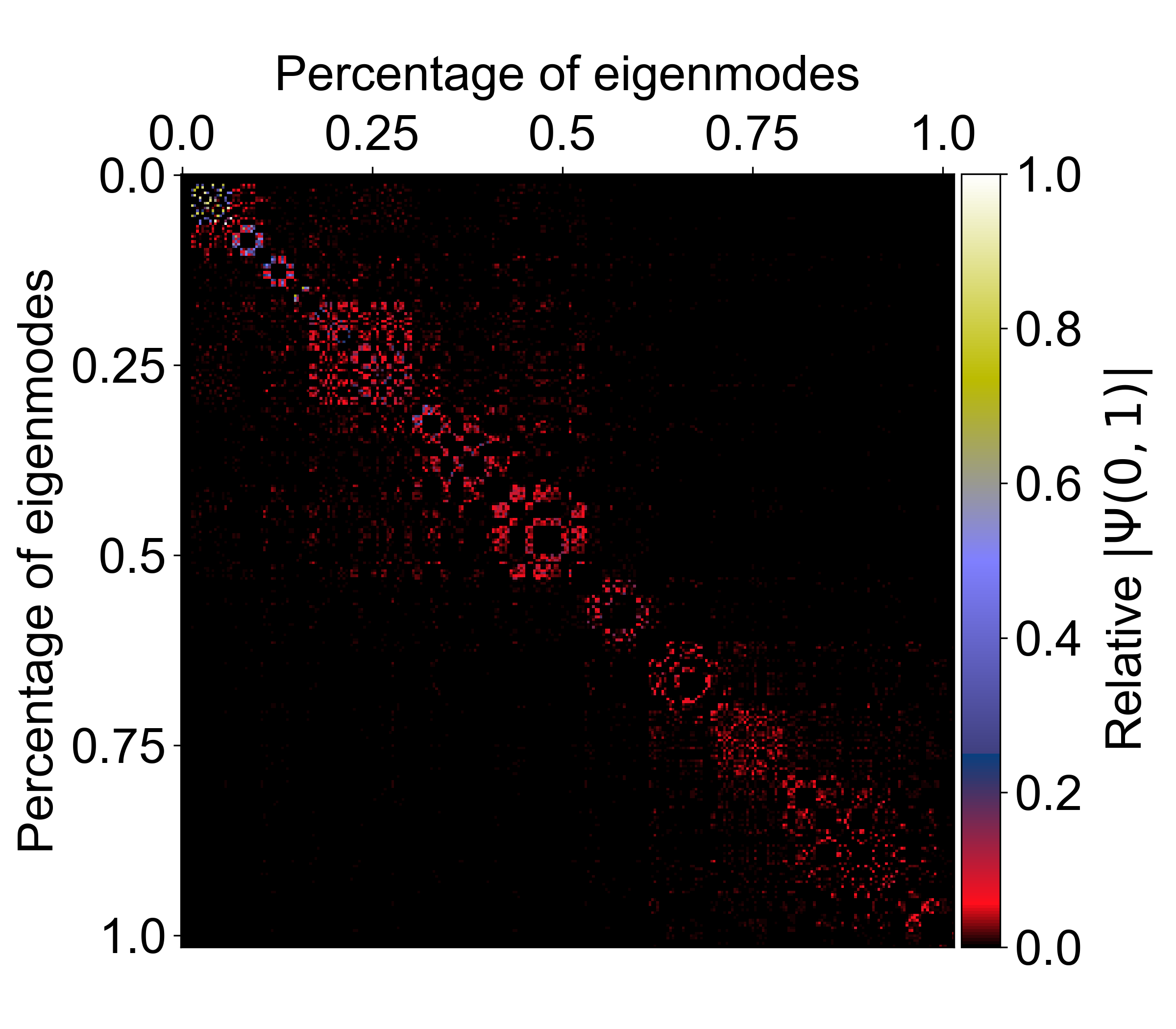}
    \caption{An intensity map of the absolute values of the intermediate matrix $\Psi$ relative to its maximum value, for the initial 1\% of its rows and columns in natural diamond at 100 K. As described in the main text, the rows and columns of $\Psi$ correspond to the eigenmodes of $\Omega$ arranged in the increasing order of their respective eigenvalues. The largest values of $\Psi$ occur along the diagonal for the eigenmodes with the smallest eigenvalues, and drop sharply away from the diagonal and along the diagonal as the eigenvalues increase in magnitude, thus motivating a rank reduction of $\Psi$ to $\Psi_R$.}
    \label{fig:intermediate_matrix}
\end{figure}

Additionally, in solving the transient LPBE, Eq.~\ref{eq:psi_matrix_element} suggests another layer of rank reduction on the intermediate matrix $\Psi$. First, the matrix elements $\Psi^{m n}$ are inversely proportional to the eigenvalues of $\Omega$. Second, the off-diagonal elements of $\Psi$ are proportional to $\mathcal{V}^{m n}$, which vanish when both $m$ and $n$ are even eigenmodes of $\Omega$. Since all of the eigenmodes of $\Omega$ with large eigenvalues are numerically even, as discussed in the main text, $\mathcal{V}^{m n}$ vanishes for large $m$ and $n$. Thus, the elements of $\Psi$ drop sharply away from the diagonal and along the diagonal, as shown in Fig.~\ref{fig:intermediate_matrix}. Therefore, a small subset block $\Psi_R$, containing a small fraction of the rows and columns of the full intermediate matrix $\Psi$, is sufficient to describe the transient phonon dynamics from Eq.~\ref{eq:psi_matrix_element}, thus significantly reducing the computational cost associated with the repeated diagonalization of $\Psi$ or $\Psi_R$ required for every $\{\xi, \eta \}$.

\section{Matrix-free diagonalization leveraging the symmetries of collision matrix} \label{sec:matrix_free_approch}
The most important step in the low-rank solution of the LPBE described in the section~\ref{sec:results} is the diagonalization of the collision matrix $\Omega$. Commonly-used solvers for matrix diagonalization, like LAPACK or ScaLAPACK, require storage of the full collision matrix as well as space for all of the calculated eigenvectors apart from the necessary workspace. Further, for ultrahigh-$\kappa$ materials, particularly at low temperatures, we require fine discretization of the Brillouin zone to resolve the weak, but necessary, momentum-dissipating Umklapp scattering processes that the thermally-occupied low frequency phonons undergo~\cite{pitaevskii_physical_2012}. This requirement translates into a significantly large memory requirement, as discussed in the Supplementary section~S3.\\

However, as elucidated in section~\ref{sec:results}, only a few eigenmodes of $\Omega$ with the smallest eigenvalues significantly contribute to heat transport. We leverage this important feature of phonon dynamics by employing the Krylov subspace diagonalization algorithm - Implicitly restarted Arnoldi method (IRAM) as implemented in the open-source package ARPACK~\cite{lehoucq_arpack_1998}, which can efficiently compute a few eigenmodes of $\Omega$ of interest. Krylov methods like IRAM rely only on the evaluation of matrix-vector products and do not require the full collision matrix, thus resulting in significant reduction in the computational memory requirement. Furthermore, the symmetry of the collision matrix $\Omega$ under the space-group operations of the crystal, i.e., $\Omega_{\left( \mathcal{S} \lambda \right) \left( \mathcal{S} \lambda_{1} \right)} = \Omega_{\lambda \lambda_{1}}$ (see Supplementary section~S2 for detailed derivation) for the crystal symmetry operation $\mathcal{S}$, facilitates further computational optimization in evaluating matrix-vector products of the form: $y = \Omega x$ to feed as input to the Krylov solvers. Here, the elements of the output vector, $y_\lambda$, which correspond to the phonon modes in the irreducible wedge of the Brillouin zone under the space-group operations~\cite{maradudin_symmetry_1968}, are obtained in the usual way:  $y_{\lambda} = \sum_{\lambda_{1}} \Omega_{\lambda \lambda_{1}} x_{\lambda_{1}}$. This operation requires only those rows of $\Omega$ corresponding to the phonon modes in the irreducible wedge. As shown in Eq.~\ref{eq:irr_matrix_vector_product}, the remaining elements of the output vector $y$ for the phonon modes outside the irreducible wedge, i.e., $y_{\mathcal{S} \lambda}$, are also computed using these irreducible rows of $\Omega$ using the symmetry transformations of $\Omega$ as:
\begin{align}
    y_{\mathcal{S} \lambda} & = \sum_{\lambda_{1}} \Omega_{\left( \mathcal{S} \lambda \right) \lambda_{1}} x_{\lambda_{1}} = \sum_{\mathcal{S} \lambda_{1}} \Omega_{\left( \mathcal{S} \lambda \right) \left( \mathcal{S} \lambda_{1} \right)} x_{\mathcal{S} \lambda_{1}} \nonumber \\
    & = \sum_{\lambda_{1}} \Omega_{\lambda \lambda_{1}} x_{\mathcal{S} \lambda_{1}}
    \label{eq:irr_matrix_vector_product}
\end{align}
Finally, we also employ level 2 BLAS routines to efficiently compute the inner products between the input vector $x$ and the irreducible rows of the collision matrix.

% \section{Author contributions}
% N.K.R. originated the research idea. N.M. developed the computational framework and performed the calculations. N.M. and N.K.R. analyzed the results and wrote the manuscript.
% \section{Data availability}
% The numerical data supporting the findings of this work will be made available by the corresponding author upon reasonable request.
% \section{Code availability}
% All formulations and computational optimizations necessary to perform the calculations presented in this manuscript are described in the Methods section, in the Supplementary information and in refs.~\cite{ravichandran_unified_2018, malviya_failure_2023}.
% \begin{acknowledgments}
% This work was supported by the Core Research Grant (CRG) No. CRG/2020/006166, and the Mathematical Research Impact Centric Support (MATRICS) grant no. MTR/2022/001043 from the Department of Science and Technology - Science and Engineering Research Board, India. NM gratefully acknowledges the Prime Minister's Research Fellowship (PMRF) grant no. PMRF-02-01036. NR thanks the Infosys Foundation for their support through a Young Investigator Award.
% \end{acknowledgments}

% \nocite{*}

\bibliography{references}% Produces the bibliography via BibTeX.

\end{document}